\documentclass{PoS}

\usepackage{graphicx}
\usepackage{amsmath}
\usepackage{amsfonts}
\usepackage{amssymb}
\usepackage{amsthm}
\usepackage{caption}
\usepackage{fontenc}
\usepackage{graphicx}
\usepackage{comment}
\usepackage{wrapfig}
\usepackage{epsfig}
\newcommand\ddfrac[2]{\frac{\displaystyle #1}{\displaystyle #2}}
\newcommand{\si}{\sigma_{eff}}
\usepackage[most]{tcolorbox}
\usepackage{textpos}
\usepackage{longfbox} 
\title{A proton imagining via double parton scattering}

\ShortTitle{A proton imagining via DPS}

\author{\speaker{M. Rinaldi}\\
        Dipartimento di Fisica e Geologia. Universit\`a degli studi di 
Perugia. INFN section of Perugia, via A. Pascoli, Perugia, Italy.\\
        E-mail: \email{matteo.rinaldi@pg.infn.it}}

\abstract{In this 
contribution  we discuss the main outcomes of our studies on
 the so called
double parton distribution functions (dPDFs),  accessible quantities in 
high
energy proton-proton and proton nucleus collisions, in double parton 
scattering 
processes (DPS). 
These new
distributions are 
 almost unknown, nevertheless  they encode  
information on how partons inside a proton are 
correlated among each
other. Double PDFs
represent a new tool to explore the three dimensional 
partonic 
structure of hadrons. 
Here,  we show results obtained from calculations of
dPFDs.  In particular we focus our attention on  the impact of double 
correlations in
 experimental observables by showing how the latter could be studied 
in the next LHC run. We also
discuss how the present knowledge on a peculiar experimental 
observable could unveil new information 
on the transverse proton structure.}

\FullConference{Light Cone 2019 - QCD on the light cone: from hadrons 
to heavy ions - LC2019\\
		16-20 September 2019\\
		Ecole Polytechnique, Palaiseau, France}

\begin{document}

\section{Introduction}
\vskip -0.2cm
 A proper description of multiple parton interactions (MPI) in 
hadron-hadron collisions, occurring when  more than one parton of a 
hadron 
interact with  partons of the other colliding hadron, is relevant 
thanks to the high luminosity reached in collider experiments, such as 
at the LHC. In addition  MPI  represent a background for the search of 
new  Physic at the 
LHC. Since,
 MPI contribution  to the total cross section is  
suppressed  with respect to the single partonic interaction,
the measurement of their cross sections  is an 
important experimental
 challenge. 
Nevertheless,
 several experimental analyses 
have been collected  for most simple case of MPI, i.e. 
the   double 
parton 
scattering (DPS)
 \cite{3a,4a,6a}. 
From a theoretical point of view, we show that new fundamental information on the 
partonic 
proton 
structure \cite{hadronic,noij3} could be accessed via DPS. 
Formally, 
the DPS cross section
depends on  the so called double 
parton distribution functions (dPDFs),
$F_{ij}(x_1,x_2,{\vec z}_\perp,\mu)$, which describe the 
joint probability 
of finding two partons of flavors $i,j=q, \bar q,g$ with 
longitudinal momentum fractions $x_1,x_2$ and distance 
$\vec z_\perp$ in the transverse plane inside the hadron \cite{1a}.
Here $\mu$ is the renormalization scale.
Since  no data are available for dPDFs,
in order to estimate the magitudo of a given DPS process, unknown double parton 
correlations are usually neglected in experimental studies.
However, 
 the latter condition must be verified.
In the present contribution we show how 
constituent quark model (CQM) calculations could be used to investigate 
the impact of double parton correlations (DPCs) in dPDFs (see  Refs. 
\cite{noij3,noi1,noij1,Courtoy:2019cxq,noice}) and  in 
experimental observables such as the so called 
$\sigma_{eff}$. 
 To this aim in Refs. 
\cite{noij1,noij2},  dPDFs have been 
studied at  
the energy 
scale of the experiments and then used to calculate $\si$   \cite{noij3,noice,noiprl,noiPLB, 
noiads}.   The DPS cross 
section, in 
processes with final 
state 
$A+B$, is written through   the following ratio \cite{MPI15}:
\vskip -0.6cm
\begin{eqnarray}
\sigma^{A+B}_{DPS}  = \dfrac{m}{2} \dfrac{\sigma_{SPS}^A 
\sigma_{SPS}^B}{\sigma_{eff}}\,,
\label{sigma_eff_exp}
\end{eqnarray}
where $m$ is combinatorial factor depending on the final states $A$ and 
$B$ 
($m=1$ for $A=B$ or $m=2$ for $A \neq B$) and
$\sigma^{A(B)}_{SPS}$ is the single parton scattering cross section 
with 
final 
state $A(B)$. 
The present knowledge on DPS cross sections
has been condensed in the experimental  extraction of 
$\sigma_{eff}$~
\cite{MPI15,S1,S2,S3,S4,data8,data9,data6,data10,data11,data12}. 
A constant value, $\sigma_{eff} \simeq$ 
15 mb, is compatible, within errors, with data: 
a
result  obtained neglecting    DPCs. In the next sections the  results of 
the calculations
$\sigma_{eff}$   within CQM will be described in
order to characterize signals of DPCs.

\subsection{Double parton correlation and the same-sign $WW$ production 
at LHC}

\begin{figure}[h]
\includegraphics{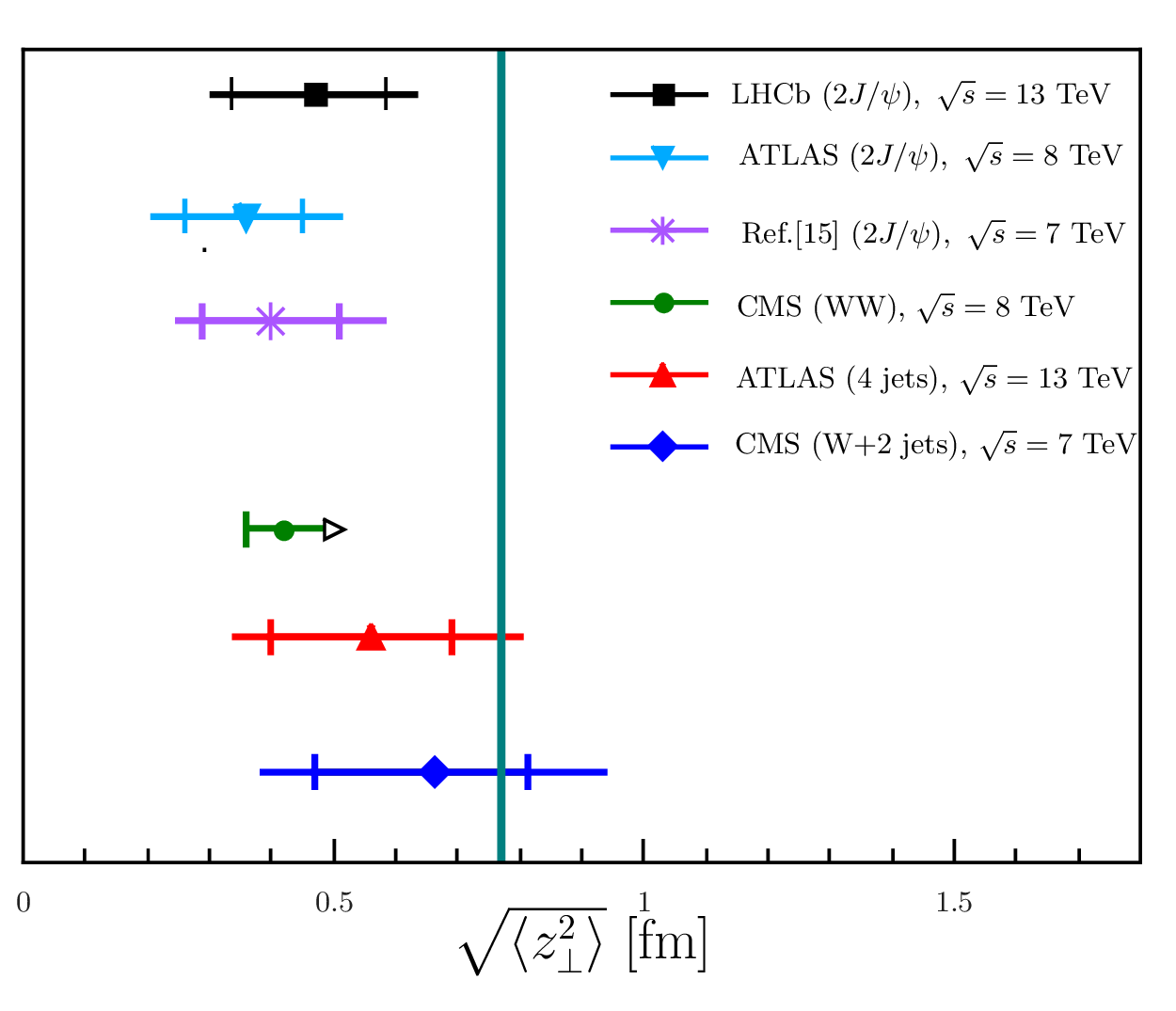}
\includegraphics{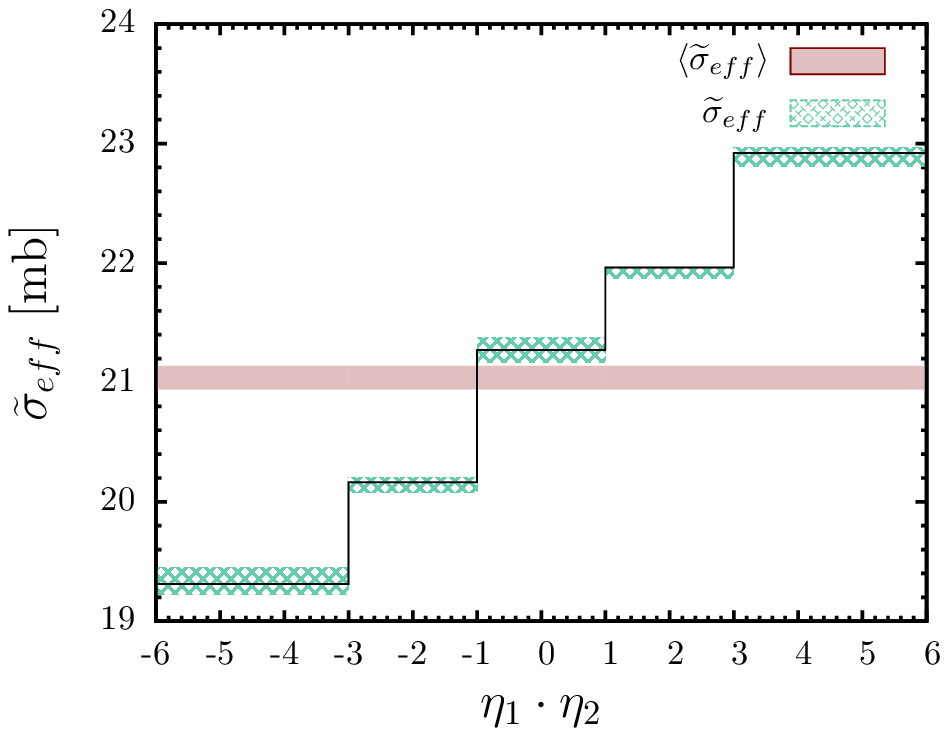}
\vskip 5.cm
\caption{\footnotesize{ 
 Left 
panel:
$\widetilde{\sigma}_{eff}$ and $\langle 
\widetilde{\sigma}_{eff} \rangle$ as a 
function of product of muon rapidities. Right panel: The application or 
Eq. (\ref{ine}) by using 
data of Refs. \cite{data8,data9,data6,data10,data11,data12}}.}
\label{efb}
\end{figure}

In Refs. \cite{noij3,noi1,noij1,noice}, non perturbative 
correlations effects in dPDFs have been 
studied. We found out that DPCs cannot be 
easily neglected also in the experimental kinematic conditions. 
In Ref. \cite{noiprl}, we have considered the  
same sign $W$ pair production process, a golden channel 
 for the observation of DPS  \cite{gaunt2,cotogno}, 
to establish to what extent double correlations could be accessed  at 
the LHC.
The differential DPS cross section can be written as follows \cite{1a}:

\vskip -0.4cm
\begin{align}
 d \sigma^{AB}_{DPS} = \dfrac{m}{2} \sum_{i,j,k,l} d \vec z_{\perp} 
F_{ij}(x_1,x_2, \vec z_\perp, \mu)F_{ij}(x_3,x_4, \vec z_\perp, \mu)
&d\hat \sigma^A_{ik} d\hat \sigma^B_{jl},
\label{dps}
\end{align}
where $\hat \sigma_{ij}^A$ represents the elementary cross section.
As non perturbative input of the calculations, use has been made of 
dPDFs evaluated in Ref. \cite{noij1}.
 Details of the fiducial DPS phase space 
adopted in the analysis and the theoretical errors are discussed in Ref. \cite{noiprl}. We 
found that  the total $W$-charge summed DPS cross section (considering 
both 
$W$ decays into same sign muons) is found to be $\sigma^{++}+\sigma^{--} 
[\mbox{fb}] \sim 
0.69$. This
result is consistent with those obtained by neglecting DPCs 
and those 
obtained with dPDFs of the model of 
Ref. 
\cite{3a}.  
The effects of DPCs have been investigated
by observing the
 $\tilde \sigma_{eff}$ dependence on   $\eta_1 \cdot \eta_2 \simeq 1/4~ 
\mbox{ln}(x_1/x_3)\mbox{ln}(x_2/x_4)$.  
For this process we found a mean value  $\langle \widetilde 
\sigma_{eff} \rangle \sim 21.04$ mb, consistently with calculations 
of Refs. \cite{noij3,noiPLB,noiads}.
A clear signature of the presence of DPCs is found by observing the
departure of   $\widetilde \sigma_{eff}$ from a constant, see right 
panel of Fig. \ref{efb}.  We have  
estimated 
that, with a luminosity of $\mathcal{L} \sim 1000 
~\mbox{fb}^{-1}$, at 68\% confidence level, the departure of 
$\sigma_{eff}$ 
from a constant value can be measured in the next run of the LHC.

\section{The  $\sigma_{eff}$: a new window towards  the transverse 
proton 
structure}
Here 
we show how the experimental estimates of $\si$ could be interpreted in 
terms of the geometric structure of the proton.
We consider the strategy used in Ref. 
\cite{hadronic}, where DPCs have neglected in order to match with the 
experimental studies.
In this scenario:
\vskip -0.5cm
\begin{align}
 \label{si}
\si^{-1} = \int d^2 z_\perp T(z_\perp)^2 = \int \frac{d^2k_\perp}{(2 
\pi)^2} \tilde T(\vec k_\perp) \tilde T(-\vec k_\perp)~,
\end{align}
where $\tilde T(k_\perp)$, called  form factor (eff), is the Fourier 
transform (FT) of the probability of finding two partons with 
transverse distance $z_\perp$, i.e. $T(z_\perp)$.
 Thanks to the probabilistic interpretation of $T(z_\perp)$ and 
the asymptotic behaviour of $\tilde T (k_\perp)$,
 $\si$ can be related
to 
the mean transverse  distance $\langle z_\perp^2 \rangle$ 
between two  partons in a DPS process \cite{hadronic,noij3}.
In 
fact $\langle 
z_\perp^2 \rangle= -4 ~d \tilde T(k_\perp)/dk_\perp^2 ~|_{k_\perp=0} $~.
In the non relativistic limit:

\vskip -0.7cm
\begin{align}
 \tilde T(k_\perp) = \int dk_1~dk_1~\Psi^\dagger(\vec k_1+\vec 
k_\perp,\vec k_2)\Psi(\vec k_1, \vec k_2+\vec k_\perp)~,
\end{align}
\vskip -0.2cm
\noindent
 where $\vec k_i$ is the momentum of a parton $i$ and $\Psi$ is the 
proton 
wave function. The above expression is similar to the 
standard proton ff. However, 
 in this case there is  a double momentum 
imbalance, i.e. $k_\perp$.
One should expect that in the extremely high 
$k_\perp$ region, the eff should fall to zero at least as the standard 
 ff, being the eff  a double ff  
\cite{hadronic,noij3}. Thanks to these  general and model 
independent 
conditions, we found:

\vskip -0.5cm
\begin{align}
\label{ine}
 \ddfrac{\si}{3 \pi} \leq \langle z_\perp^2 \rangle \leq 
\ddfrac{\si}{\pi}~.
\end{align}
\vskip -0.cm
The above expression has been verified by using all models of dPDFs at 
our disposal, even for the pion target. In the right panel of Fig. 
\ref{efb} the 
experimental values of $\si$  have 
been  used in Eq. (\ref{ine}) to get the mean 
transverse  distance of the active partons.
The above relation has been properly generalised in Ref. \cite{noij3} 
in order to include correlations  
and splitting effects.
Thanks to these analyses, new information on the transverse structure 
of the proton can 
be obtained from detailed experimental analyses of $\si$.

\section{Conclusions}
In this contribution we have 
shown our main outcomes about dPDFs, remarking the impact of double parton correlations in 
experimental observables 
 such as $\si$. We found that the 
dependence of $\si$ upon the longitudinal momentum fraction of the 
partons represents a clear sign correlations. We estimated that their effects could be observed in the 
next LHC run.
From a different perspective, we have shown that new 
information on the mean transverse distance of partons inside the 
proton could be obtained from measurements of $\si$. 
All these studies 
suggest that
 dPDFs are fundamental 
tools to obtain new details on the non perturbative structure of the 
proton.

\subsection{Acknowledgments}
The author thanks all the orginzers of the conference for the support 
given for 
this talk. The author also thanks F. A. Ceccopieri, S. Scopetta, M. 
Traini and V. Vento for their collaboration to this talk. 
This  work  was  sup-ported in part by the STRONG-2020 project of the 
European Union's Horizon 2020 research and innovation programme under 
grant agreement No 824093.

\bibliography{iopart-num2}

\end{document}